\documentclass[journal=jctcce,layout=onecolumn,manuscript=letter]{achemso}
\setkeys{acs}{doi = true}
\newcommand{\doi}[1]{\href{http://dx.doi.org/#1}{\nolinkurl{#1}}}

\usepackage{amssymb}
\usepackage{amsmath}
\usepackage{amsfonts}
\usepackage{booktabs}
\usepackage{siunitx}
\usepackage[hidelinks]{hyperref}
\usepackage{mathrsfs}
\usepackage{mathtools}
\usepackage{multirow}
\usepackage{pdflscape}
\usepackage{physics}
\usepackage{subfig}
\usepackage{pdfpages}

\newcommand{\spqe}[1]{SPQE}
\newcommand{\adaptsd}[1]{ADAPT-VQE-SD}
\newcommand{\adaptgsd}[1]{ADAPT-VQE-GSD}

\title{Linear-Scaling Quantum Circuits for Computational Chemistry}

\author{Ilias Magoulas}
\email{ilias.magoulas@emory.edu}
\affiliation
{Department of Chemistry and Cherry Emerson Center for Scientific Computation,
	Emory University, Atlanta, Georgia 30322, USA}

\author{Francesco A.\ Evangelista}
\email{francesco.evangelista@emory.edu}
\affiliation
{Department of Chemistry and Cherry Emerson Center for Scientific Computation,
	Emory University, Atlanta, Georgia 30322, USA}

\date{\today}

\begin{document}
	
	\begin{abstract}
		
		We have recently constructed compact, CNOT-efficient, quantum circuits for fermionic and 
		qubit excitations of arbitrary many-body rank [I.\ Magoulas and F.A.\ Evangelista, 
		\textit{J.\ Chem.\ Theory Comput.}\ \textbf{19}, 822 (2023)]. Here, we present 
		approximations to these circuits that substantially reduce the CNOT counts even further. 
		Our preliminary numerical data, using the selected projective quantum eigensolver
		approach,
		demonstrate that there is practically no loss of accuracy in 
		the energies compared to the parent implementation while the ensuing symmetry breaking is 
		essentially negligible.
		
	\end{abstract}
	
	\maketitle
	
	Chemistry has been identified as one of the first potential killer applications for quantum 
	computing.\cite{Reiher.2017.10.1073/pnas.1619152114} This is due to the fact that a quantum 
	device can simulate a chemical problem with a number of computer elements (qubits) that
	scales, in principle, linearly rather than exponentially with system size. Even if an
	exponential advantage cannot be achieved for every chemical problem of 
	interest,\cite{Lee.2022.2208.02199} any form of polynomial speed up could potentially 
	bring classically intractable applications within computational reach.
	
	Several low-depth hybrid quantum--classical approaches have been proposed that are suitable for 
	current noisy intermediate-scale quantum hardware. In general, they can be divided into two 
	broad categories. The first family contains algorithms that rely on an ansatz, such as the 
	variational (VQE),\cite{Peruzzo.2014.10.1038/ncomms5213,McClean.2016.10.1088/1367-2630/18/2/023023,
		Cerezo.2021.10.1038/s42254-021-00348-9,Tilly.2021.2111.05176,Fedorov.2022.10.1186/s41313-021-00032-6}
	contracted,\cite{Smart.2021.10.1103/PhysRevLett.126.070504} and
	projective\cite{Stair.2021.10.1103/PRXQuantum.2.030301} (PQE) quantum eigensolvers, while 
	the second is comprised of ansatz-independent schemes, including quantum imaginary time 
	evolution\cite{Motta.2020.10.1038/s41567-019-0704-4,Sun.2021.10.1103/PRXQuantum.2.010317} and 
	quantum subspace diagonalization methods.\cite{Motta.2020.10.1038/s41567-019-0704-4,
		McClean.2017.10.1103/PhysRevA.95.042308,Parrish.2019.1909.08925v1,Stair.2020.10.1021/acs.jctc.9b01125,
		Huggins.2020.10.1088/1367-2630/ab867b}
	
	Focusing on ansatz-dependent techniques that interest us more for the purposes of this 
	work, the trial state is expressed in terms of a unitary parameterization, \latin{i.e.},
	\begin{equation} 
	\ket*{\tilde{\Psi}(\mathbf{t})} = U(\mathbf{t}) \ket*{\Phi},
	\end{equation}
	where $\mathbf{t}$ denotes a set of parameters and $\ket*{\Phi}$ is a reference state that can 
	be easily prepared on the quantum device, usually the Hartree--Fock Slater determinant. 
	Chemically inspired ans\"{a}tze are almost invariably based on the unitary 
	extension\cite{Kutzelnigg.1977.10.1007/978-1-4757-0887-5_5,Kutzelnigg.1982.10.1063/1.444231,
		Kutzelnigg.1983.10.1063/1.446313,Kutzelnigg.1984.10.1063/1.446736,Bartlett.1989.10.1016/S0009-2614(89)87372-5,
		Szalay.1995.10.1063/1.469641,Taube.2006.10.1002/qua.21198,Cooper.2010.10.1063/1.3520564,
		Evangelista.2011.10.1063/1.3598471,Harsha.2018.10.1063/1.5011033,Filip.2020.10.1063/5.0026141,
		Freericks.2022.10.3390/sym14030494,Anand.2022.10.1039/d1cs00932j}
	of coupled-cluster theory\cite{Coester.1958.10.1016/0029-5582(58)90280-3,
		Coester.1960.10.1016/0029-5582(60)90140-1,Cizek.1966.10.1063/1.1727484,Cizek.1969.10.1002/9780470143599.ch2,
		Cizek.1971.10.1002/qua.560050402,Paldus.1972.10.1103/PhysRevA.5.50}
	(UCC). In general, a factorized form of the UCC unitary is adopted,
	\begin{equation}
		\label{disentangled_ucc}
		U(\mathbf{t}) = \prod_{\mu} e^{t_\mu \kappa_\mu},
	\end{equation}
	also known as disentangled UCC,\cite{Evangelista.2019.10.1063/1.5133059} that can be readily 
	implemented on a quantum device. The $\kappa_\mu$ symbols appearing in eq 
	\eqref{disentangled_ucc} represent generic fermionic, anti-Hermitian, particle--hole excitation
	operators. For an $n$-tuple excitation, they are defined as
	\begin{equation}
	\kappa_\mu \equiv 
	\kappa_{i_1 \ldots i_n}^{a_1 \ldots a_n} = a^{a_1} \cdots a^{a_n} a_{i_n} \cdots a_{i_1} - a^{i_1} 
	\cdots a^{i_n} a_{a_n} \cdots a_{a_1},
	\end{equation}
	where $a_p$ ($a^p \equiv a_p^\dagger$) is the second-quantized
	annihilation (creation) operator acting on spinorbital $\phi_p$ and indices $i_1, i_2, \ldots$ or
	$i, j, \dots$ ($a_1, a_2, \ldots$ or $a, b, \ldots$) label spinorbitals occupied (unoccupied) in $\ket*{\Phi}$.  
	An alternative strategy that leads to more efficient quantum circuits is to replace the 
	fermionic anti-Hermitian operators by their qubit counterparts, defined as
	\begin{equation}
		Q_{i_1 \ldots i_n}^{a_1 \ldots a_n} = Q^{a_1} \cdots Q^{a_n} Q_{i_n} \cdots Q_{i_1} - Q^{i_1} 
		\cdots Q^{i_n} Q_{a_n} \cdots Q_{a_1},
	\end{equation}
	with $Q_p$ ($Q^p \equiv Q_p^\dagger$) denoting the qubit annihilation (creation) operator acting
	on the $p^\text{th}$ qubit. However, in doing so, one may 
	potentially sacrifice the proper sign structure of the resulting
	state,\cite{Yordanov.2020.10.1103/PhysRevA.102.062612,Yordanov.2021.10.1038/s42005-021-00730-0,
		Xia.2021.10.1088/2058-9565/abbc74,Mazziotti.2021.10.1088/1367-2630/ac3573,
		Smart.2022.10.1103/PhysRevA.105.062424,Magoulas.2023.10.1021/acs.jctc.2c01016}
	since qubit excitations neglect the fermionic sign.
	Designing efficient, \latin{i.e.}, low-depth and noise-resilient, quantum circuits representing 
	fermionic and qubit excitations is crucial for the success of ansatz-dependent algorithms on 
	current noisy quantum hardware.
	
	Inspired by the work of Yordanov \latin{et 
	al.},\cite{Yordanov.2020.10.1103/PhysRevA.102.062612,Yordanov.2021.10.1038/s42005-021-00730-0} 
	we have recently introduced compact fermionic- (FEB) and qubit-excitation-based 
	(QEB) quantum circuits that efficiently implement excitations of arbitrary many-body 
	rank\cite{Magoulas.2023.10.1021/acs.jctc.2c01016} (see, also, ref \citenum{Xu.2023.2302.08679}
	for an alternative CNOT-efficient approach that requires ancilla qubits).
	While the FEB/QEB quantum circuits are equivalent to their 
	conventional analogs, \latin{i.e.}, there is no loss of accuracy in the simulations, they 
	significantly reduce the numbers of single-qubit and, more importantly, CNOT gates (recall that 
	experimental realizations of two-qubit gates, such as CNOT, tend to have errors that are about 10 
	times larger than those of the single-qubit ones\cite{google_weber}). For example, the standard
	quantum circuit implementing a hextuple qubit excitation 
	requires more than 45,000 CNOT gates while its QEB counterpart needs only about 2,000. Despite
	the drastic reduction in the CNOT count afforded by the FEB and QEB 
	quantum circuits, their number continues to scale exponentially with the operator many-body rank. 
	Consequently, quantum algorithms based on a full FEB/QEB operator pool will typically generate circuits 
	with unfavorable CNOT counts when compared to approaches relying on pools 
	containing lower-rank excitation operators, such as singles and doubles or their generalized 
	extension.\cite{Nooijen.2000.10.1103/PhysRevLett.84.2108,Nakatsuji.2000.10.1063/1.1287275}
	
	As elaborated on in our earlier study,\cite{Magoulas.2023.10.1021/acs.jctc.2c01016} the
	multi-qubit-controlled $R_y$ gate is the dominant source of CNOTs in the FEB/QEB
	quantum circuits. In that work, we relied on an ancilla-free implementation of that gate that requires
	$2^{2n-1}$ CNOTs, where $n$ is the many-body rank of the given excitation
	operator. Adopting more efficient implementations of the multiply controlled $R_y$ gate can significantly
	reduce the CNOT requirements. For
	example, the approach advocated in ref \citenum{Maslov.2016.10.1103/PhysRevA.93.022311} results
	in the linear-scaling CNOT count of $12n-14$, but requires $\lceil (2n - 3)/2 \rceil$
	ancilla qubits, where $\lceil x \rceil$ denotes the ceiling of $x$. Recently, ancilla-free,
	CNOT-efficient implementations of multiply controlled gates have been proposed. Of particular
	interest are the ones introduced in refs \citenum{daSilva.2022.10.1103/PhysRevA.106.042602} and
		\citenum{Vale.2023.2302.06377},
	which decompose the multi-qubit-controlled $R_y$ gate into circuits containing $16n^2 - 24n + 10$
	and, at most, $32n - 40$ CNOTs, respectively. All of these state-of-the-art decompositions
	generate FEB/QEB quantum circuits with significantly less CNOT gates compared to those reported
	in our earlier study, especially so as the many-body rank increases. Nevertheless, they either
	require ancilla qubits, have a $\mathscr{O}(n^2)$ scaling, or have large prefactors in the resulting
	CNOT counts. 

	\begin{figure*}
		\centering
		\captionsetup[subfigure]{labelformat=empty}
		\subfloat[]{\includegraphics[width=0.65\linewidth]{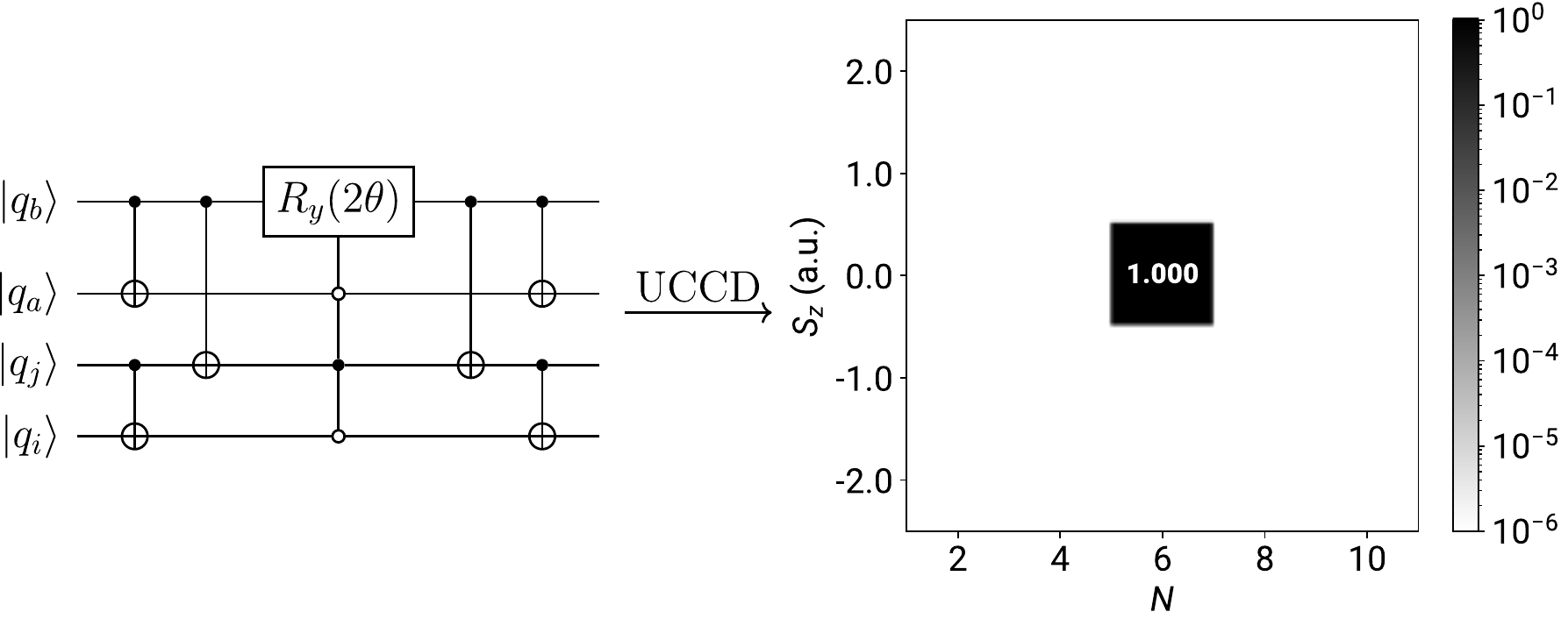}} \\\vspace*{-12pt}
		\subfloat[]{\includegraphics[width=0.65\linewidth]{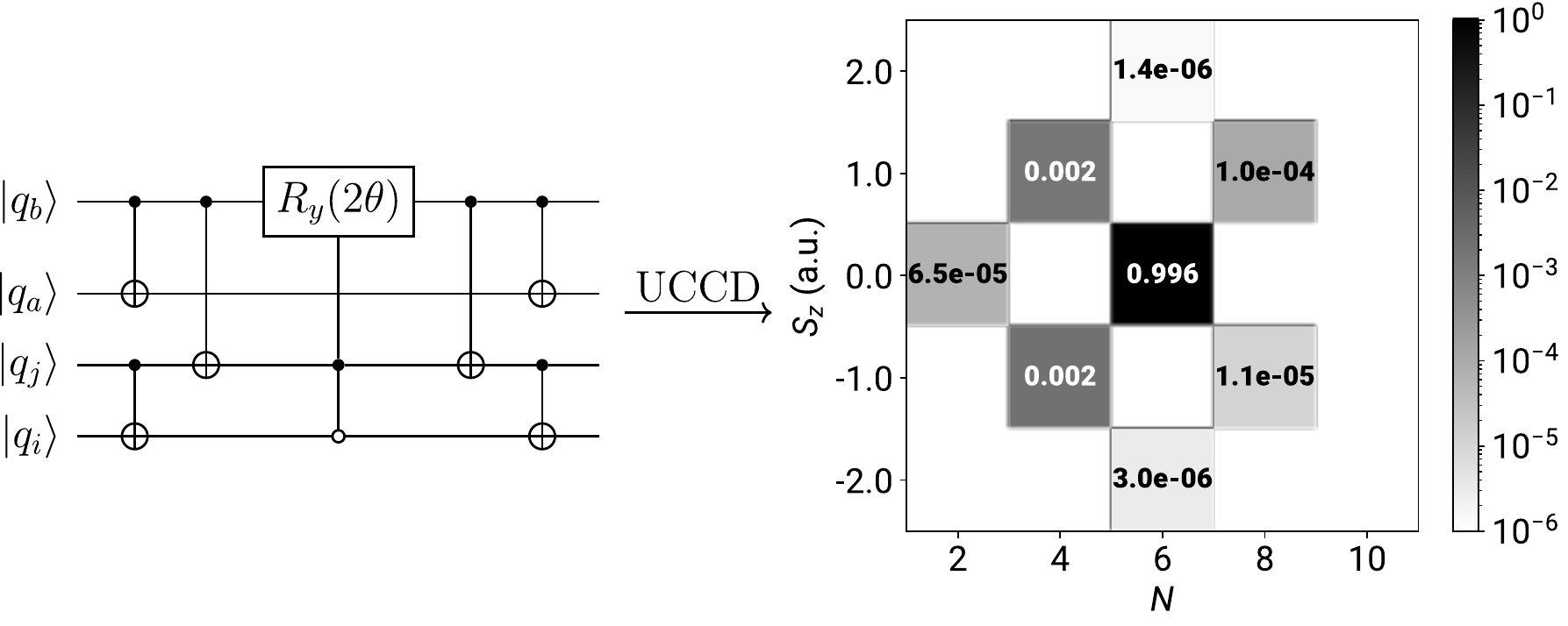}} \\\vspace*{-12pt}
		\subfloat[]{\includegraphics[width=0.65\linewidth]{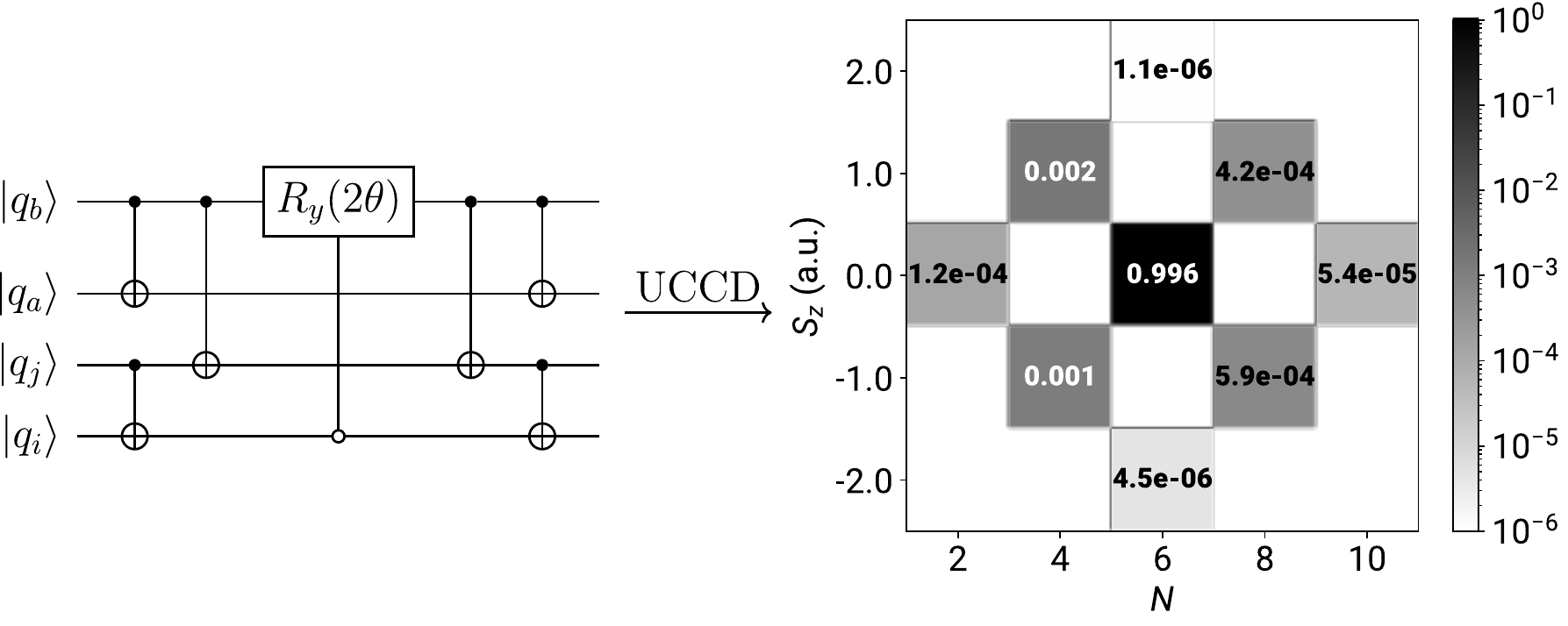}} \\\vspace*{-12pt}
		\subfloat[]{\includegraphics[width=0.65\linewidth]{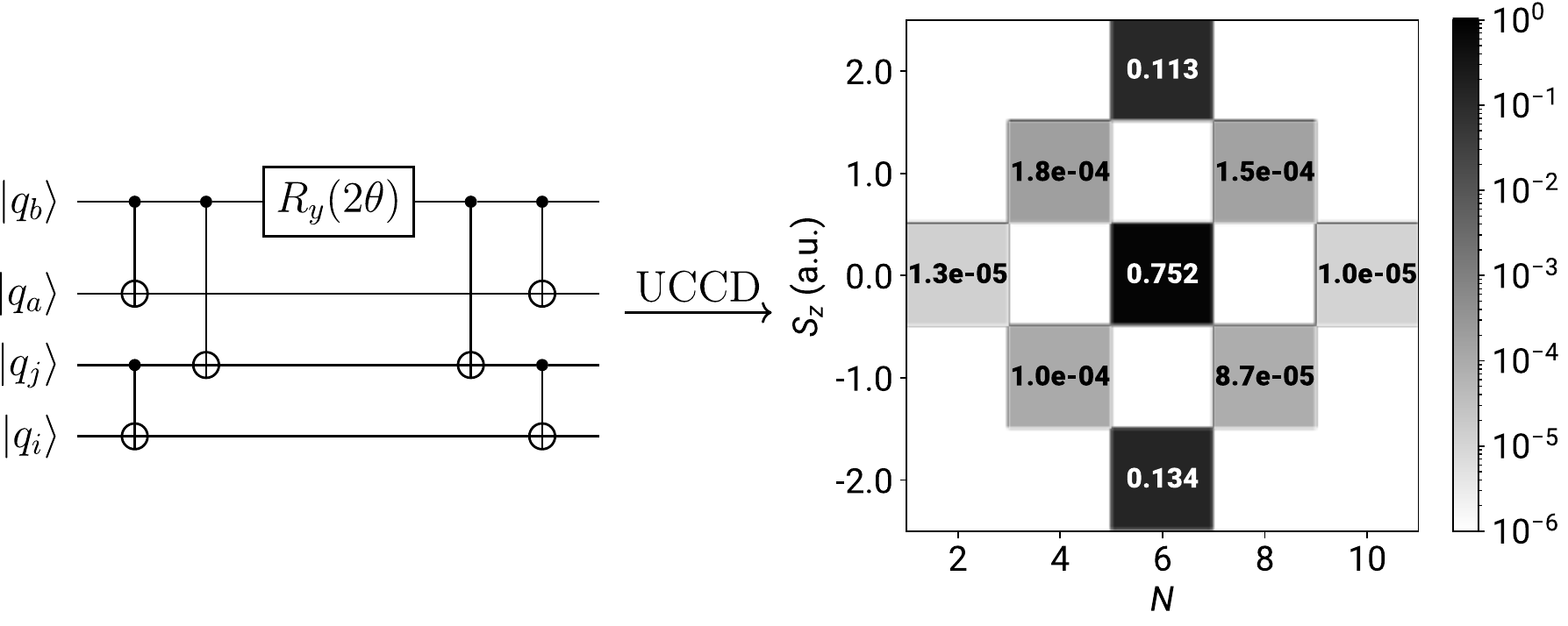}}\vspace*{-12pt}
		\caption{\label{figure1}
			Illustration of the $N$- and $S_z$-symmetry breaking introduced by
			the removal of controls from the multiply controlled $R_y$ gate appearing
			in the FEB/QEB quantum circuits. On the left we give the
			relevant qubit double excitation circuits and on the right we provide the
			contribution of each symmetry sector of the Fock space to the converged
			wavefunctions. Since spatial symmetry is conserved, only the totally symmetric
			part of the Fock space is considered. The depicted data resulted from VQE QEB-UCCD
			simulations of the $\text{H}_6$/STO-6G linear chain with a separation
			between neighboring H atoms of \SI{2.0}{\AA}.
		}
	\end{figure*}
	In our efforts to design CNOT-frugal FEB/QEB quantum circuits, we opted for a different strategy.
	In this letter, we consider approximate implementations of the 
	multi-qubit-controlled $R_y$ gate in which the number of control qubits is reduced. 
	Since the resulting circuits are not equivalent to their parent FEB/QEB counterparts, some loss 
	of accuracy in the computed energies is anticipated. Furthermore, as shown analytically in the 
	Supporting Information, the removal of control qubits leads to the breaking of the 
	particle number ($N$) and total spin projection ($S_z$) symmetries, while spatial
	symmetry is still preserved. To demonstrate this effect, we performed single-point, VQE UCC with doubles
	(UCCD) simulations using the full QEB circuits and three approximations, the numerical results of which
	are depicted in Figure \ref{figure1}. In these illustrative calculations, we focused on the $\text{H}_6$
	linear chain, as described by the STO-6G minimum basis.\cite{Hehre.1969.10.1063/1.1672392}
	The geometry that we selected was characterized by the distance between neighboring hydrogen atoms
	($R_{\text{H--H}}$) of \SI{2.0}{\AA}, the largest H--H separation  considered in our earlier 
	study.\cite{Magoulas.2023.10.1021/acs.jctc.2c01016}
	As shown in Figure \ref{figure1}, the removal of controls from the multi-qubit-controlled $R_y$
	gate leads to a `leaking' of the wavefunction into other symmetry sectors of the Fock space. Specifically, 
	we observe contaminants with eigenvalues of $N$ and $S_z$ that differ by $\pm 2$ and $\pm 4$ for
	$N$, and $\pm 1$ and $\pm \SI{2}{{a.u.}}$ for $S_z$, relative to the $N = 6$
	and $S_z = \SI{0}{{a.u.}}$ values characterizing the ground electronic state of $\text{H}_6$. This observation is consistent with
	the analytical results presented in the Supporting Information. As might have been anticipated, we find that
	the more controls are removed, the more severe the symmetry breaking becomes, as illustrated in Figure \ref{figure1}.
	
	Consequently,
	the guiding principle in designing such approximate FEB/QEB quantum circuits has been to find 
	a good compromise between reducing the CNOT count and minimizing the loss of accuracy in the 
	computed energies and the breaking of symmetries in the final states. In the Supporting 
	Information, we consider various approximate schemes, implemented in a local version of the
	QForte package.\cite{Stair.2022.10.1021/acs.jctc.1c01155} We performed single-point selected 
	PQE\cite{Stair.2021.10.1103/PRXQuantum.2.030301} (SPQE) simulations 
	for the challenging $\text{H}_6$/STO-6G linear chain with $R_{\text{H--H}} = \SI{2.0}{\AA}$.
	Recall that the SPQE algorithm typically relies on a complete pool of particle--hole excitation
	operators to iteratively construct the ansatz, eq \eqref{disentangled_ucc}, and the optimum 
	parameters are obtained by enforcing the residual condition
	\begin{equation}
		r_\mu \equiv \mel{\Phi_\mu}{U^\dagger(\mathbf{t}) H U(\mathbf{t})}{\Phi} = 0
	\end{equation}
	for all excited Slater determinants $\ket*{\Phi_\mu}$ corresponding to the excitation
	operators $\kappa_\mu$ appearing in the ansatz unitary $U(\mathbf{t})$ (the details of
	the PQE and SPQE approaches can be found in refs \citenum{Stair.2021.10.1103/PRXQuantum.2.030301}
	and \citenum{Magoulas.2023.10.1021/acs.jctc.2c01016}).
	Based on these preliminary computations, the best balance is offered by the following recipe 
	(see the Supporting Information for the details):
	\begin{itemize}
		\item Single and double excitations are treated fully [see panels (a) and (b) of Figure S6].
		\item For triple and quadruple excitations, only controls over qubits corresponding to 
		occupied spinorbitals are retained in the multi-qubit-controlled $R_y$ gate [see panels (c)
		and (d) of Figure S6].
		\item For pentuple and higher-rank excitations, all controls are removed [see Figure S6(e)],
		\latin{i.e.}, the multi-qubit-controlled $R_y$ gate is replaced by its single-qubit
		analog. 
	\end{itemize}
	In the case of higher-rank excitation operators, the above procedure reduces the scaling of
	the CNOT count with the excitation rank from exponential to linear. For qubit excitations,
	in particular, the number of CNOT gates becomes $4n-2$, where $n$ is the excitation rank.
	
	To assess the effectiveness of the above approximation scheme, denoted as aFEB for 
	fermionic and aQEB for qubit excitations, and to compare it with the parent 
	FEB/QEB quantum circuits across a wide range of correlation effects, we 
	performed SPQE simulations of the symmetric dissociation of the $\text{H}_6$/STO-6G linear 
	chain. The grid of H--H distances used to sample the potential energy curve (PEC) was 
	$R_\text{H--H} = 0.5, 0.6, \ldots, \SI{4.0}{\AA}$. All SPQE simulations reported in this work 
	utilized a full operator pool and micro- and macro-iteration thresholds of 
	\SI{e-5}{\textit{E}_h} and \SI{e-2}{\textit{E}_h}, respectively (see refs 
	\citenum{Stair.2021.10.1103/PRXQuantum.2.030301} and \citenum{Magoulas.2023.10.1021/acs.jctc.2c01016}
	for the details of the recently proposed SPQE algorithm). To ensure a lower number of residual element 
	evaluations, the PQE micro-iterations employed the direct inversion of the iterative 
	subspace\cite{Pulay.1980.10.1016/0009-2614(80)80396-4,Pulay.1982.10.1002/jcc.540030413,
		Scuseria.1986.10.1016/0009-2614(86)80461-4}
	(DIIS) accelerator and the maximum number of micro-iterations was set to 50. All correlated 
	approaches were based on restricted Hartree--Fock references with the one- and two-electron 
	integrals obtained from Psi4.\cite{Smith.2020.10.1063/5.0006002}
	
\begin{figure}[!h]
		\centering
		\includegraphics[width=0.3\textwidth]{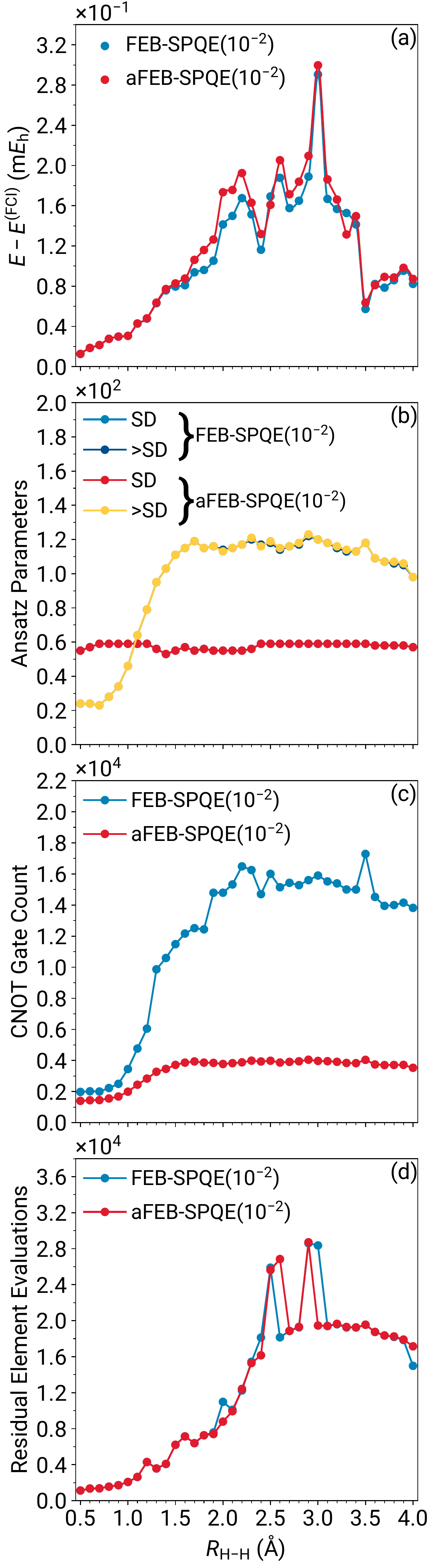}
		\caption{\label{figure2}
			Errors relative to FCI [(a)], ansatz parameters [(b)], CNOT gate counts 
			[(c)], and residual element evaluations [(d)] characterizing the FEB- and 
			aFEB-\spqe{2} simulations of the symmetric dissociation of the linear
			$\text{H}_6$/STO-6G system. The ``SD'' and ``\textgreater 
			SD'' 
			symbols in the legend to panel (b) denote single or double (SD) or higher (\textgreater 
			SD) 
			excitation operators.
		}
	\end{figure}

	We begin the discussion of our numerical results by examining the ability of the aFEB-SPQE 
	approach to reproduce the parent FEB-SPQE simulations and reduce the required
	computational resources. To that end, in Figure \ref{figure2}, we compare the energies, 
	numbers of operators in the converged ansatz unitaries, CNOT counts, and numbers of residual 
	element evaluations obtained with FEB-\spqe{2} and aFEB-\spqe{2}, 
	characterizing the symmetric dissociation of the $\text{H}_6$/STO-6G linear chain. A quick 
	inspection of Figure \ref{figure2} immediately reveals that aFEB-\spqe{2} is both a highly 
	accurate approximation to FEB-\spqe{2} and computationally efficient. In the case
	of energetics, aFEB-\spqe{2} faithfully reproduces the data of the full 
	FEB-\spqe{2} approach, being characterized by mean absolute, maximum absolute, and non-parallelity 
	error values of 10, 32, and \SI{53}{\micro \textit{E}_h}, respectively. As far as the 
	computational resources are concerned, aFEB-\spqe{2} captures 
	practically the identical numbers of parameters when compared to FEB-\spqe{2} [see panel (b) of 
	Figure \ref{figure2}]. Nevertheless, as illustrated in Figure \ref{figure2}(c), aFEB-\spqe{2} 
	generates quantum circuits with significantly reduced numbers of CNOT gates than full 
	FEB-\spqe{2}. As might have been anticipated from the nature of the approximation, the 
	disparity between the aFEB- and FEB-\spqe{2} CNOT counts is dramatically increased as 
	the strength of non-dynamic correlations increases, with aFEB-SPQE requiring up to 4 times less 
	CNOTs than its full FEB counterpart. Finally, 
	the aFEB- and FEB-\spqe{2} schemes 
	require more or less the same numbers of residual element evaluations.
	Consequently, despite the drastic nature of the approximation in the quantum circuits, 
	aFEB-\spqe{2} accurately reproduces the FEB-\spqe{2} energies and, by extension, those of full 
	configuration interaction (FCI), but at a tiny fraction of the computational cost of its 
	FEB-\spqe{2} parent. This observation is true for the entire range of electron correlation effects
	characterizing the symmetric dissociation of the $\text{H}_6$/STO-6G linear chain.
	
	\begin{figure}[!h]
		\centering
		\includegraphics[width=0.3\textwidth]{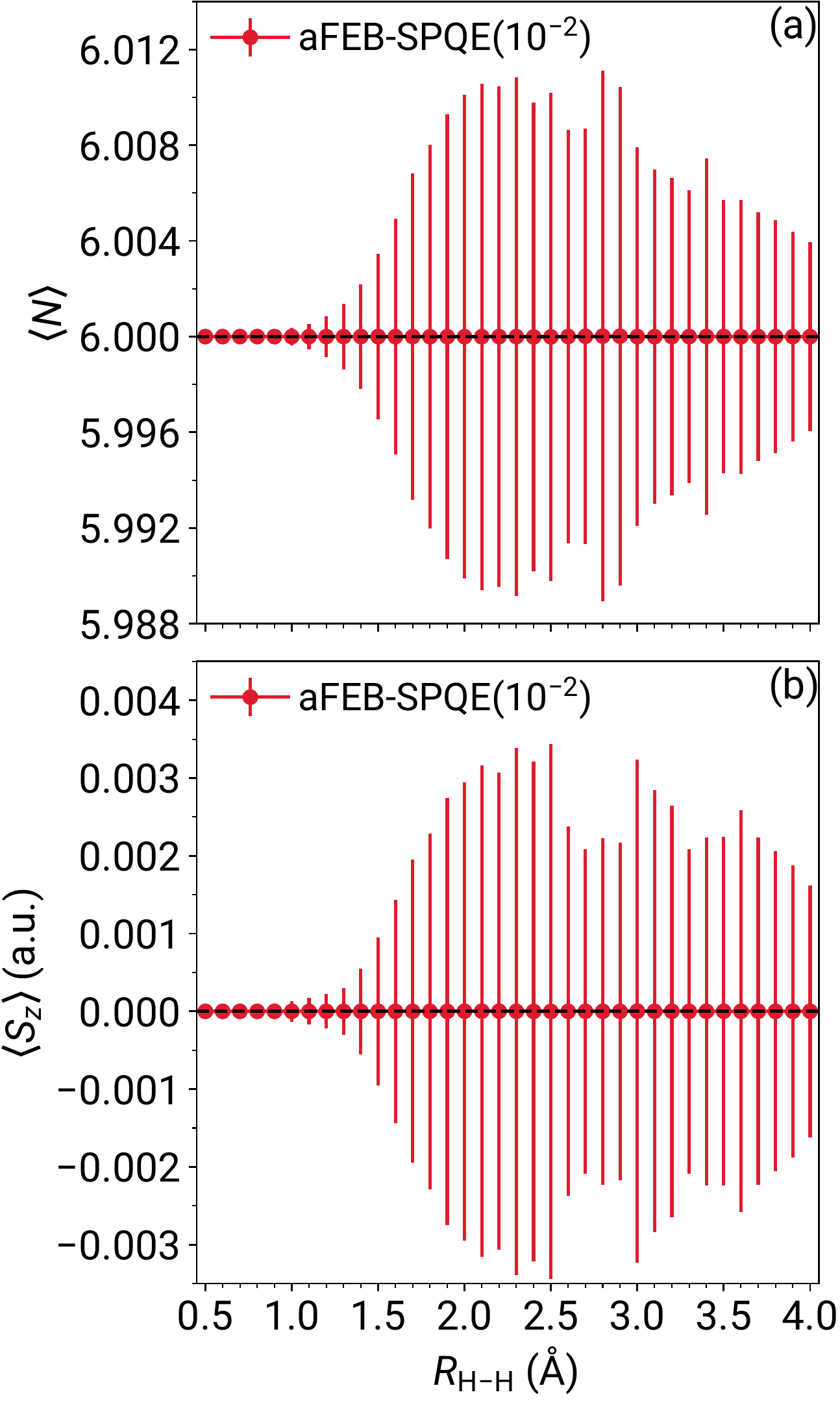}
		\caption{\label{figure3}
			Expectation values of (a) the particle number $N$ and (b) the projection of the total 
			spin on the $z$ axis $S_z$ operators characterizing the aFEB-\spqe{2} simulations of 
			the symmetric dissociation of the linear $\text{H}_6$/STO-6G system. The vertical lines 
			denote standard deviations, computed as $\sigma_A = \sqrt{\ev*{A^2} - \ev*{A}^2}$.
			The horizontal dashed lines denote the corresponding eigenvalues for the ground 
			electronic state of the $\text{H}_6$/STO-6G linear chain.
		}
	\end{figure}
	Despite the excellent performance in recovering the FEB-\spqe{2} energetics, as already 
	mentioned above and elaborated on in the Supporting Information, the approximations in the 
	underlying quantum circuits defining the aFEB-\spqe{2} approach result in the breaking of the 
	particle number $N$ and total spin projection $S_z$ symmetries. It is, thus, worth examining the 
	degree to which these symmetries are broken. As illustrated in Figure \ref{figure3}, the 
	expectation values of the $N$ and $S_z$ operators are essentially identical to the eigenvalues 
	of 6 and \SI{0}{{a.u.}}, respectively, characterizing the ground electronic state of the linear 
	$\text{H}_6$ system. Indeed, the maximum unsigned errors are $2 \times 10^{-5}$ in the 
	case of $N$ and \SI{3e-6}{{a.u.}} for $S_z$. However, due to the fact that the symmetry 
	breaking introduces contaminants with both lower and higher eigenvalues of $N$ and $S_z$, 
	expectation values are not a good metric. By examining the error bars shown in Figure 
	\ref{figure3}, given by the standard deviation $\sigma_A = \sqrt{\ev*{A^2} - \ev*{A}^2}$,
	the following trend becomes apparent. In the weakly 
	correlated regime, there is practically no symmetry breaking. As all H--H distances are symmetrically 
	stretched, the standard deviations gradually increase in the recoupling region until they reach their 
	maximum values, around $R_\text{H--H} = \SI{2.5}{\AA}$. Finally, as $\text{H}_6$ approaches its dissociation
	limit, the errors gradually decrease. This pattern directly correlates with the number of 
	higher-than-double excitation operators in the ansatz, shown in Figure \ref{figure2}(b). This 
	behavior is not surprising since the aFEB approximate scheme relies on a full implementation 
	of singles and doubles, \latin{i.e.}, the higher-than-double excitation operators are the sole 
	source of $N$- and $S_z$-symmetry contaminants. 
	The maximum standard deviations of $\max(\sigma_N)= 0.011$ and 
	$\max(\sigma_{S_z}) = \SI{0.003}{{a.u.}}$ are, respectively, two and three orders of magnitude smaller than 
	the distance of 1 between the neighboring eigenvalues of $N$ and $S_z$. This observation provides
	further evidence supporting the notion that the aFEB scheme induces negligible symmetry breaking
	effects. As a definitive proof, we computed the weight of the totally symmetric Slater determinants
	with $N = 6$ and $S_z = 0$ in the final wavefunctions. Focusing on the $R_\text{H--H} = \SI{2.5}{\AA}$
	and \SI{2.8}{\AA} geometries, corresponding to $\max(\sigma_{S_z})$ and $\max(\sigma_N)$, respectively,
	we find that the weight of determinants having the correct symmetry properties are 99.998\% and
	99.999\%.

	\begin{figure}[!h]
		\centering
		\includegraphics[width=0.3\textwidth]{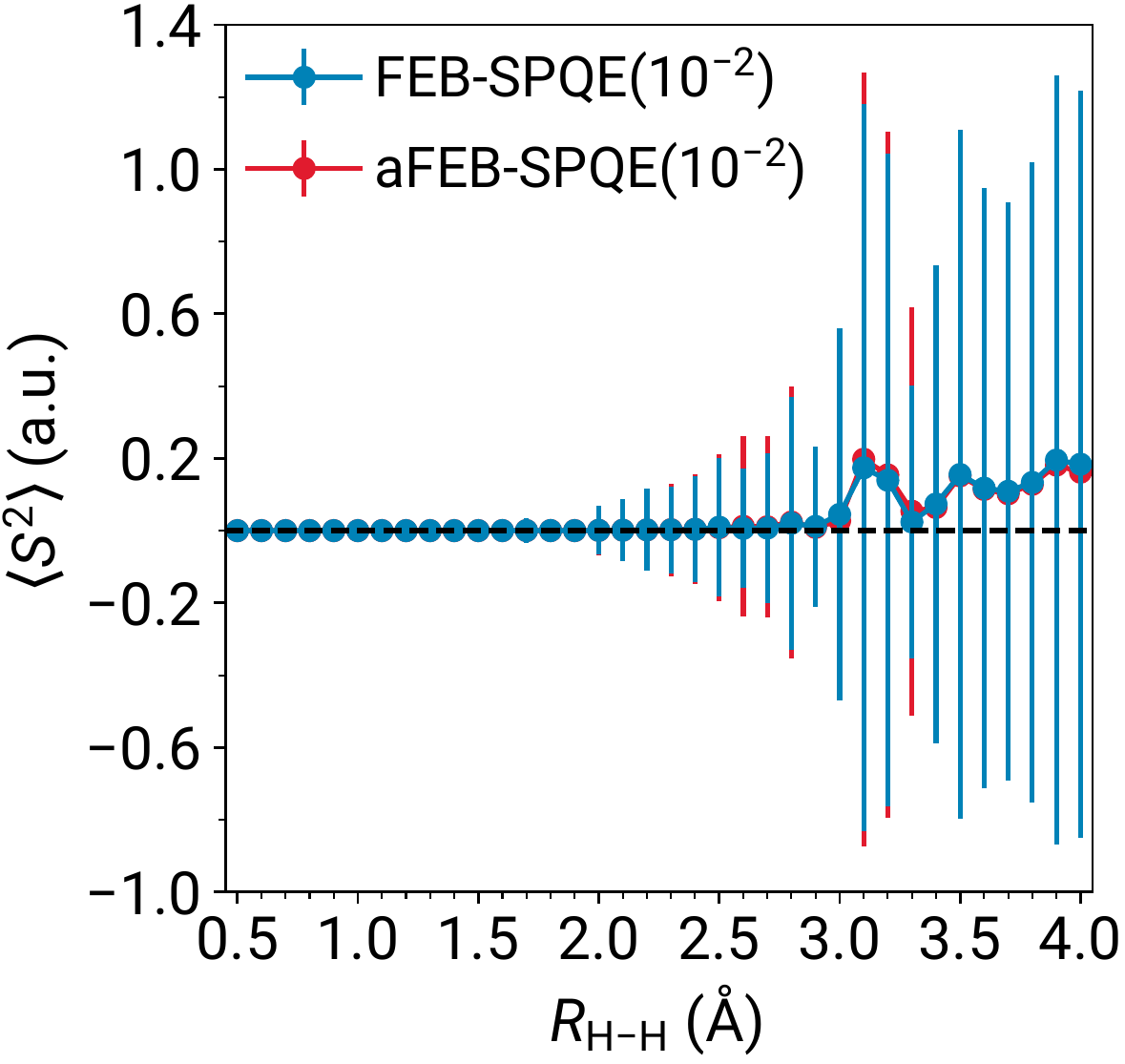}
		\caption{\label{figure4}
			Expectation values of the total spin squared $S^2$ operator characterizing the FEB- and 
			aFEB-\spqe{2} simulations of the symmetric dissociation of the linear 
			$\text{H}_6$/STO-6G system. The vertical lines denote standard deviations, computed as 
			$\sigma_A = \sqrt{\ev*{A^2} - \ev*{A}^2}$. The horizontal dashed line denotes the 
			corresponding eigenvalue for the ground electronic state of the $\text{H}_6$/STO-6G 
			linear chain.
		}
	\end{figure}	
	Due to the use of a determinantal basis, the converged states resulting from FEB- and 
	aFEB-\spqe{2} simulations are not necessarily eigenfunctions of the square of the total spin 
	operator, $S^2$. Nevertheless, it is still interesting to examine how the $\ev*{S^2}$ and 
	$\sigma_{S^2}$ values are affected when one transitions from the parent FEB-\spqe{2} scheme to 
	its aFEB approximation. As depicted in Figure \ref{figure4}, aFEB-\spqe{2} yields nearly
	identical $\ev*{S^2}$ and $\sigma_{S^2}$ values with those obtained with the full FEB-\spqe{2}
	approach. This further reinforces the fact that aFEB-\spqe{2} is a high-fidelity approximation
	to FEB-\spqe{2}.
	
	Although here we focused on the aFEB-/FEB-\spqe{2} pair, as shown in Figures S7--S9, 
	similar observations can be made when examining the performance of the aQEB approximation to 
	QEB-\spqe{2}. In comparing the two approximate schemes among themselves (Figures S10 and S11),
	we notice that 
	aQEB-\spqe{2} typically produces quantum circuits with fewer CNOT gates than its fermionic 
	counterpart, especially in situations characterized by stronger non-dynamic correlation 
	effects. At the same time, however, aQEB-\spqe{2} is typically less accurate than aFEB-\spqe{2} 
	and the symmetry breaking is more pronounced. These observations indicate that
	aFEB-\spqe{2} achieves a favorable balance between minimizing the CNOT count and mitigating the
	loss of accuracy in energetics and symmetry breaking in the final states.
	
	
	Our preliminary numerical results advocate that the aFEB scheme has several desirable 
	properties of an approximation. It is highly accurate, reproducing the parent FEB-\spqe{2} 
	simulations with errors not exceeding a few microhartree. It has a low computational cost, 
	reducing the number of CNOT gates compared to its already efficient FEB analog by 65\%, on 
	average. Furthermore, the aFEB quantum circuits are much simpler compared to their FEB 
	counterparts, suggesting an easier hardware implementation. One aspect of aFEB-SPQE that we 
	intend to examine in the future is its stability. Although preliminary single-point 
	calculations for the $\text{H}_6$ ring, the $\text{H}_8$ linear chain, the linear 
	$\text{BeH}_2$ system, and the $C_{\text{2v}}$-symmetric insertion of Be to $\text{H}_2$ 
	indicate that aFEB-\spqe{2} behaves similarly to the case of the $\text{H}_6$ linear chain, a 
	more thorough investigation is required. It is also worth exploring the usefulness of symmetry 
	restoration\cite{Tsuchimochi.2020.10.1103/PhysRevResearch.2.043142,Khamoshi.2021.10.1088/2058-9565/abc1bb,
	Khamoshi.2023.10.1088/2058-9565/ac93ae} within the various approximations considered in this work.
	As shown in our preliminary single-point calculations reported in Table S1, restoring
	the $N$ and $S_z$ symmetries in aFEB-/aQEB-SPQE has a negligible effect in the computed energies. This 
	is due to the fact that the symmetry breaking in these approximations is practically 
	insignificant. Nevertheless, symmetry restoration might prove useful in the context of more 
	drastic approximations. In such cases, it might be possible to reduce the CNOT counts even 
	further while still maintaining a high degree of accuracy in the computed energies.
	

	\begin{acknowledgement}
		
		This work is supported by the U.S.\ Department of Energy under Award No.\ DE-SC0019374.
		
	\end{acknowledgement}

	\begin{suppinfo}
	
		Details on the approximate fermionic- and qubit-excitation-based quantum circuits,
		analysis of the ensuing symmetry breaking, results of additional numerical simulations. 
		
		\noindent Numerical data generated in this study.
		
	\end{suppinfo}
	
	
	\bibliography{refs}

\includepdf[pages=-,pagecommand={},width=\textwidth]{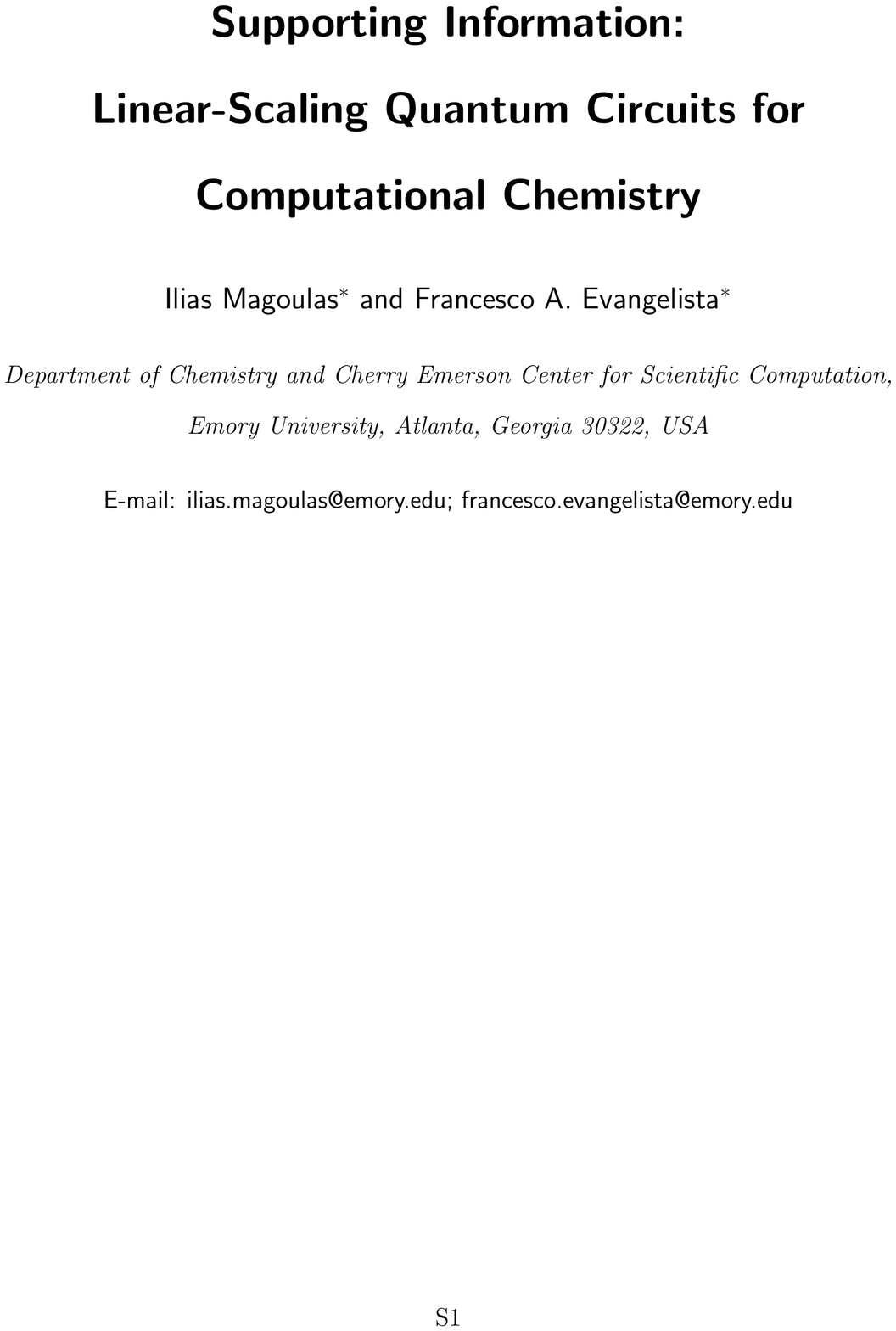}	
	
\end{document}